\documentclass[a4paper,12pt]{article}

\usepackage{color}
\definecolor{rosso}{cmyk}{0,1,1,0.4}
\definecolor{rossos}{cmyk}{0,1,1,0.55}
\definecolor{rossoc}{cmyk}{0,1,1,0.2}
\definecolor{blu}{cmyk}{1,1,0,0.3}
\definecolor{blus}{cmyk}{1,1,0,0.6}

\usepackage[dvipsnames]{xcolor} 
\usepackage{lscape}
\usepackage{colortbl}
\usepackage{multirow} 
\usepackage{tabularx}
\usepackage{pbox}
\usepackage{hhline}
\usepackage{epstopdf}

\usepackage{url}
\usepackage{amsmath}
\usepackage[final]{graphicx}
\usepackage{caption}
\usepackage{subcaption}
\usepackage{bm}
\usepackage{amssymb}

\begin{document}
\hfill
\begin{minipage}{20ex}\small
ZTF-EP-17-03\\
\end{minipage}

\begin{center}
{\LARGE \bf 
Extended Higgs Sectors\\ 
in Radiative Neutrino Models\\} 
\vspace{0.3in}
{\bf 
Oleg Antipin$^{a}$, 
Petar~\v{C}uljak$^{b}$,
Kre\v{s}imir~Kumeri\v{c}ki$^{b}$  
and Ivica~Picek$^{b}$
}\\[4ex]
{\sl
$^{a}$Rudjer Bo\v{s}kovi\'c Institute, Division of Theoretical Physics,
Bijeni\v{c}ka 54, HR-10000 Zagreb, Croatia\\[1.5ex]
$^{b}$Department of Physics, Faculty of Science, University of Zagreb,
 P.O.B. 331, HR-10002 Zagreb, Croatia\\}

\end{center}

\vspace{0.2in}
\begin{center}\large\bf Abstract\end{center}

Testable Higgs partners may be sought within the extensions of the SM Higgs sector aimed at generating neutrino masses at the loop level. We study a viability of extended Higgs sectors for two selected models of radiative neutrino masses:
a one-loop mass model, providing the Higgs partner within a real triplet scalar representation, and a three-loop mass model, providing it within its two-Higgs-doublet sector. 
The Higgs sector in the one-loop model may remain stable and perturbative
up to the Planck scale, whereas the three-loop model calls for a UV completion
around $10^6$ GeV. Additional vector-like lepton and exotic scalar fields,
which are required to close one- and three-loop neutrino-mass diagrams,
play a decisive role for the testability of the respective models. 
We constrain the parameter space of these models using LHC bounds on diboson resonances.

\vspace{0.2in}

\begin{flushleft}
\small
\emph{Keywords}: Extensions of Higgs sector; Neutrino mass; Collider phenomenology
\end{flushleft}

\clearpage

\section{Introduction}

After discovery of a Higgs-like $125$ GeV particle~\cite{Aad:2012tfa,Chatrchyan:2012ufa}, 
the present data allow it to be just a detected part of an extended Standard Model (SM) Higgs sector. Additional Higgs 
candidates have been proposed in extensions of the SM Higgs sector aimed at generating neutrino masses at the loop level.
We study a possible appearance of a Higgs partner in the context of beyond-SM (BSM) fields which appear in two different 
models of radiative neutrino masses, displayed in Table~\ref{models}:\\ 
{\em The one-loop neutrino mass model}~\cite{Brdar:2013iea} with minimal BSM representations, providing the neutral component of a real scalar field $\Delta$  in the adjoint representation of the $SU(2)_L$ as the Higgs relative;\\ 
{\em The three-loop neutrino mass model}~\cite{Culjak:2015qja} with exotic BSM representations, where the Higgs partner
emerges in the form of the heavy CP-even neutral scalar field in the framework of the two-Higgs-doublet  model (2HDM)~\cite{Branco:2011iw}.\\ 
\begin{table}[ht]
\footnotesize
\begin{center}
\begin{tabularx}{\textwidth}{| *{4}{>{\centering\arraybackslash} c|} | *{4}{>{\centering\arraybackslash} c|}}
 \hhline{----||----}
 \multicolumn{4}{|c||}{\raisebox{-0.5ex}{\includegraphics[width=0.45\textwidth]{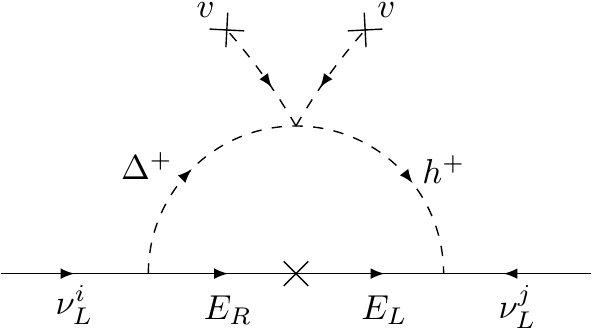}}} & \multicolumn{4}{c|}{\raisebox{-0.5ex}{\includegraphics[width=0.45\textwidth]{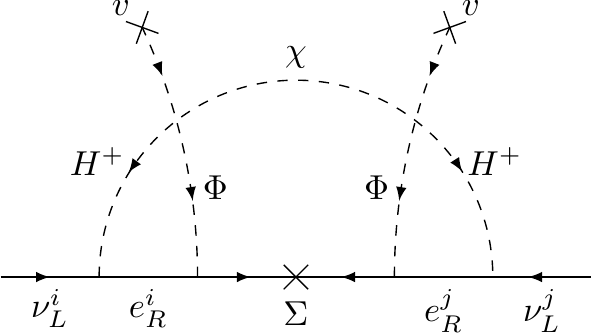}}} \\
 \hhline{----||----}
 \rowcolor[cmyk]{0.2,0.0,0,0.0}  Name &  $SU(2)_L$ &  $U(1)_Y$ & $Q$ &  Name &  $SU(2)_L$ &  $U(1)_Y$ &  $Q$ \\
 \hhline{----||----}
 \rowcolor[cmyk]{0,0,0.2,0} $\textcolor{LimeGreen}{\boldsymbol{\Delta}}$ & $3$ & $0$ & $\pm1,0$ & $\textcolor{LimeGreen}{\boldsymbol{H}}_{\textcolor{LimeGreen}{\boldsymbol{1,2}}}$ & $2$ & $1$  & $0,1$ \\
 \hhline{----||----}
  \rowcolor[cmyk]{0,0,0.2,0} $h^+$ & $1$ & $2$  & $1$ & $ \Phi$ & $5$  & $-2$ &$-3,-2,\pm 1,0$ \\
 \hhline{----||----}
 \rowcolor[cmyk]{0,0.2,0,0.1}$E_{R}$ & $2$ & $-1$ & $0,-1$ & \cellcolor[cmyk]{0,0,0.2,0} $\chi$ & \cellcolor[cmyk]{0,0,0.2,0} $7$ & \cellcolor[cmyk]{0,0,0.2,0} $0$  & \cellcolor[cmyk]{0,0,0.2,0} $\pm3,\pm2,\pm1,0$\\
 \hhline{----||----}
 \rowcolor[cmyk]{0,0.2,0,0.1}$E_{L}$ & $2$ & $-1$  & $0,-1$ & $\Sigma $ & $5$ & $0$ & $\pm2,\pm1,0$\\
 \hhline{----||----}
\end{tabularx}
\end{center}
\caption{ \rm  Neutrino mass models. Scalar fields are in (light) yellow and fermion fields in three generations are in (dark) red. The fields containing an extra Higgs candidate are in (light grey) green. For the one-loop model (left) the SM Higgs doublet  $H_{\rm SM}=(\phi^+,\phi^0)^T$ manifests itself only via its VEV $v$ in the neutrino mass diagram.}
\label{models}
\end{table}
Since the extra scalar states affect the ultra-violet (UV) behavior of a given model, such states cannot come alone or with arbitrary coupling. Previous accounts~\cite{Ferreira:2015rha} and~\cite{Khan:2016sxm} studied a validity up to Planck scale for sole second Higgs doublet or additional Higgs triplet, respectively. The present study considers such extra scalars in the setup~\cite{Brdar:2013iea,Culjak:2015qja} decorated 
with additional fields required to close respective neutrino mass loop diagrams.
The presence of extra vector-like leptons makes both radiative models
to belong to generic setup~\cite{Dermisek:2016via} which enables identification of regions of model parameter space for which the
particular cascade decays may reveal the Higgs partners.

The paper is structured as follows. In Sec.~\ref{Review} we briefly review these radiative neutrino mass models and in Sec.~\ref{diphoton} study their implications for the Higgs-partner decay signals. 
We discuss the stability of the scalar potential as well as Landau poles of relevant couplings in Sec.~\ref{perturbativity}
and present our conclusions in Sec.~\ref{Conclusions}.

\section{Two radiative neutrino mass models} \label{Review}

{\bf The one-loop model:} The first mass model~\cite{Brdar:2013iea} in our focus (LHS in Table~\ref{models}) has an appeal to invoke low non-singlet weak representations and to be free of imposing additional {\it ad hoc} $Z_2$ symmetry to eliminate the tree-level contribution. This allows a mixing of the triplet scalar field with the SM Higgs field  and its participation in electroweak symmetry breaking (EWSB).
This model has been invented on a line of attempts~\cite{Law:2013dya} to find a substitute for the ruled out one-loop model by Zee~\cite{Zee:1980ai}. The charged scalar singlet $h^+ \sim (1,2)$ from Zee's loop-diagram has been kept, while Zee's second Higgs doublet has been replaced by hypercharge zero triplet scalar field $\Delta  \sim (3,0)$
in conjunction with the vector-like lepton $E_{R,L} \sim (2,-1)$ in three generations.
The gauge invariant scalar potential contains new quartic terms
\begin{eqnarray}\label{potential}
    &&V(H_{\rm SM},\Delta,h^+)  \supset
    \lambda_3 (\mathrm{Tr}[\Delta^2])^2 + \lambda_4 H_{\rm SM}^\dag H_{\rm SM} h^-h^+ + \lambda_5 H_{\rm SM}^\dag H_{\rm SM} \mathrm{Tr}[\Delta^2] \nonumber\\
    &&+ \lambda_6 h^-h^+ \mathrm{Tr}[\Delta^2] + (\lambda_7 H_{\rm SM}^\dag\Delta\tilde{H}_{\rm SM}h^+ + \mathrm{H.c.}) + \mu H_{\rm SM}^\dag \Delta H_{\rm SM} \ ,
\end{eqnarray}
and the trilinear $\mu$ term. 
Without imposing $Z_2$ symmetry  the $\mu$ term
leads to an induced VEV  $\langle \Delta^0\rangle$ for the neutral triplet component, which is constrained by electroweak measurements to be smaller than a few GeV.
Assuming the mass values  $m_E \sim m_{\Delta^+} \sim m_{h^+} \sim 400$ GeV, the neutrino masses $m_\nu \sim 0.1$ eV can be generated for appropriate Yukawa couplings and for $\lambda_7$ coupling in (\ref{potential}) of the order of $10^{-4}$~\cite{Brdar:2013iea}. 

{\bf The three-loop model:} The second mass model~\cite{Culjak:2015qja} in our focus is based on the three-loop diagram displayed on the RHS in Table~\ref{models}.
Notably, this model contains 2HDM sector augmented by exotic fields needed to close the three-loop mass diagram: the complex scalar pentuplet $\Phi$ and a real scalar septuplet $\chi$, in conjunction with vector-like lepton $\Sigma_{R,L} \sim (5,0)$. Since $\Phi$ and $\chi$ fields do not form renormalizable gauge invariant couplings with SM fermions, there is no need for an additional symmetry to eliminate the tree-level neutrino mass contributions.
Moreover, the model is fortuitously scotogenic~\cite{Culjak:2015qja}:
a standard discrete $\tilde Z_2$ symmetry imposed to produce a natural flavour conservation in 2HDM results in accidental $Z_2$ odd parity of its BSM sector shown in Table~\ref{Z-charges}.
On account of it, the lightest among the three generations  ($\alpha =1,2,3$) of exotic real fermions $\Sigma_{\alpha} \sim (5,0)$ may be a viable minimal dark matter (MDM)~\cite{Cirelli:2005uq} candidate. 
Out of four different ways the Higgs doublets are 
conventionally assigned charges under a $\tilde Z_2$ symmetry~\cite{Kanemura:2014bqa}, we adopt 
the  ``lepton-specific" (Type X or Type IV) 2HDM implemented originally in~\cite{Aoki:2008av, Aoki:2009vf}. The ratio of VEVs of 2HDM fields $H_{\bf{1,2}}\sim (2,1)$ is given
by $\mathrm{tan}\beta = v_1/v_2$.
\begin{table}
\begin{center}
  \begin{tabular}{|c|ccccc|cc|ccc|}
   \hline
 \hbox{Symmetry} & $Q_i$ & $u_{i R}$ & $d_{i R}$ & $L_{i L}$ & $e_{i R}$ & $H_{\bf{1}}$ & $H_{\bf{2}}$ & $\Phi$ &
    $\chi$ & $\Sigma_{\alpha}$ \\\hline
\rowcolor[cmyk]{0.1,0,0.1,0}
\cellcolor[cmyk]{0.0,0,0.2,0}$Z_2\frac{}{}$                {\rm accidental} & $+$ & $+$ & $+$ & $+$ & $+$ & $+$ & $+$ & $-$ & $-$ & $-$ \\ \hline  
 \rowcolor[cmyk]{0.2,0.0,0,0.0}
  \cellcolor[cmyk]{0,0.2,0.0,0}
  $\tilde{Z}_2\frac{}{}$ {\rm imposed}& $+$ & $-$ & $-$ & $+$ &
                       $+$ & $+$ & $-$ & $+$ & $-$ & $+$ \\\hline
   \end{tabular}
\end{center}
  \caption{Charge assignment under an automatic $Z_2$ symmetry which is induced by the imposed $\tilde Z_2$ symmetry 
  in the lepton-specific 2HDM.}
  \label{Z-charges}
\end{table}
The fields $H_{\bf{1,2}}$ are expressed in standard way in terms of 
physical charged scalars $H^\pm$ and two CP-even neutral states  $h$ and $H$ which mix with the angle $\alpha$, and are proposed to be the SM-like Higgs $h(125)$ and its heavier relative $H$.

The most general CP-conserving 2HDM potential $V(H_{\bf{1}},H_{\bf{2}})$ presented in~\cite{Culjak:2015qja} is conventionally expressed in terms of five quartic couplings $\lambda_1$ to $\lambda_5$ which can be traded for the four physical Higgs boson masses and the mixing parameter $\sin(\beta-\alpha)$.
The full scalar potential contains additional gauge invariant pieces for additional exotic scalar fields $\Phi \sim (5,-2)$ and $\chi \sim (7,0)$
\begin{eqnarray}\label{scalarpot}
 V(H_{\bf{1}},H_{\bf{2}},\Phi, \chi) &=& V(H_{\bf{1}},H_{\bf{2}}) + V(\Phi)  + V(\chi) + V_m(\Phi,\chi) \\ 
\nonumber   &+&  V_m(H_{\bf{1}},H_{\bf{2}},\Phi) + V_m(H_{\bf{1}},H_{\bf{2}},\chi)
            +  V_m(H_{\bf{1}},H_{\bf{2}},\Phi, \chi)  \ .
\end{eqnarray}
The  EWSB in $\tilde Z_2$-symmetric quartic term 
  $V_m(H_{\bf{1}},H_{\bf{2}},\Phi, \chi) =  \kappa H_{\bf{1}} H_{\bf{2}} \Phi \chi + \mathrm{h.c.}$
leads to the substitution
$\kappa (H^{+}_{\bf{1}} H^{0}_{\bf{2}} +  H^{+}_{\bf{2}} H^{0}_{\bf{1}})  \rightarrow  
v \kappa \cos(2\beta) H^{+}$.
The resulting three-loop diagram~\cite{Culjak:2015qja} produces small neutrino masses with non-suppressed couplings: the values of $\mathcal{O}(1)$ for the quartic $\kappa$ and the appropriate Yukawa couplings easily reproduce neutrino masses $m_\nu \sim 0.1$ eV.

\section{Testability of extended Higgs sectors} \label{diphoton}

{\bf The one-loop model:} Since the hyperchargeless scalar triplet participates in EWSB, let us discuss how testable it is. 
After EWSB, the neutral components of the SM Higgs doublet $\phi^0$ and the triplet $\Delta^0$ mix with an angle $\theta_0$, yielding $h(125)$ and its heavier partner $H$.
Thereby the VEV for the neutral triplet component is constrained by the electroweak measurements to be $\langle \Delta^0\rangle <\mathcal{O}$(1) GeV and we neglect effects of $\mathcal{O}(\langle \Delta^0\rangle/v)$. Accordingly, the contributions of charged scalars $\Delta^\pm$ and $h^+$ to the one-loop-generated $H \to \gamma \gamma$ decay are negligible since
the leading trilinear couplings $\lambda_{H \Delta^+ \Delta^-}\sim  \lambda_3 \langle \Delta^0\rangle $ and $\lambda_{H h^+ h^-}\sim  \lambda_6 \langle \Delta^0\rangle $ vanish in the limit $\langle \Delta_0\rangle=0$, and the effects of the remaining quartic couplings are negligible.
We now turn our attention to the quartics which are associated with charged components $\Delta^\pm$ of the triplet and the charged scalar singlet $h^+$, entering into quantum loops for the production and decays of light  SM-like Higgs $h(125)$. 

\begin{figure}
\centerline{\includegraphics[scale=0.68]{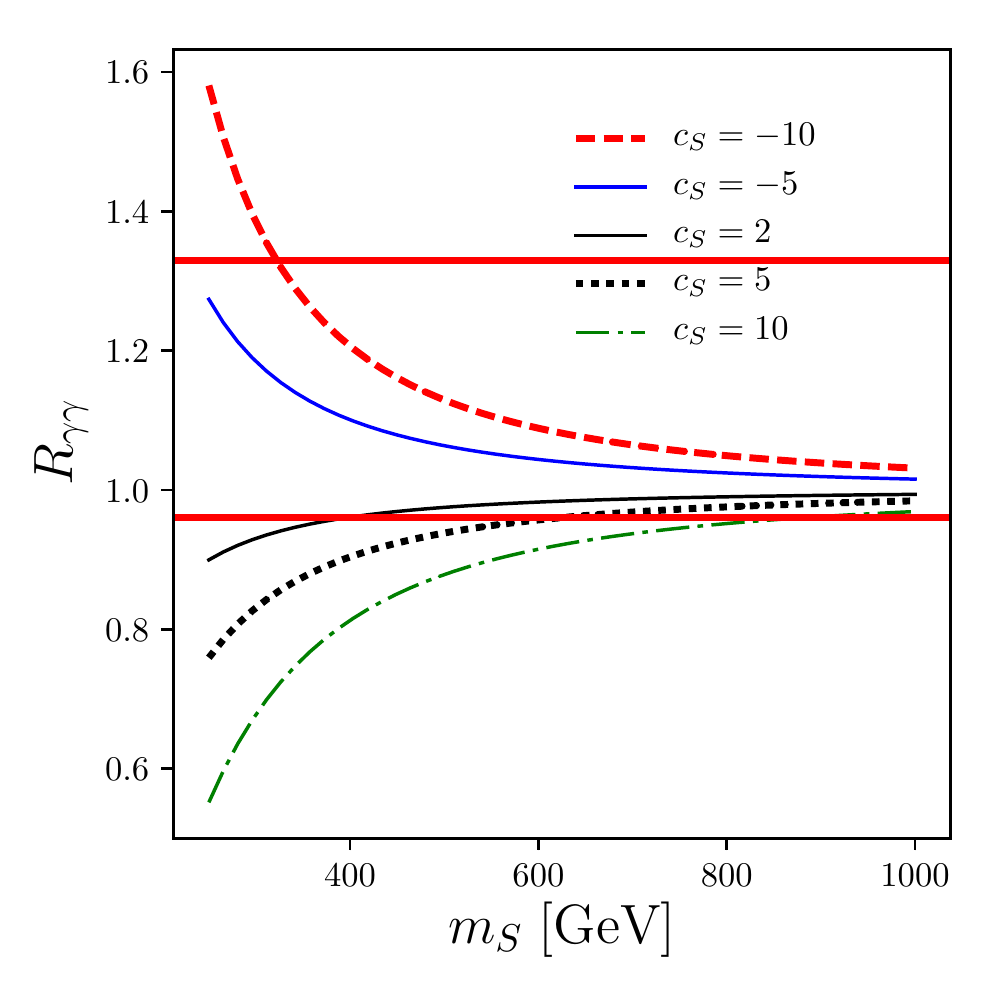}\includegraphics[scale=0.68]{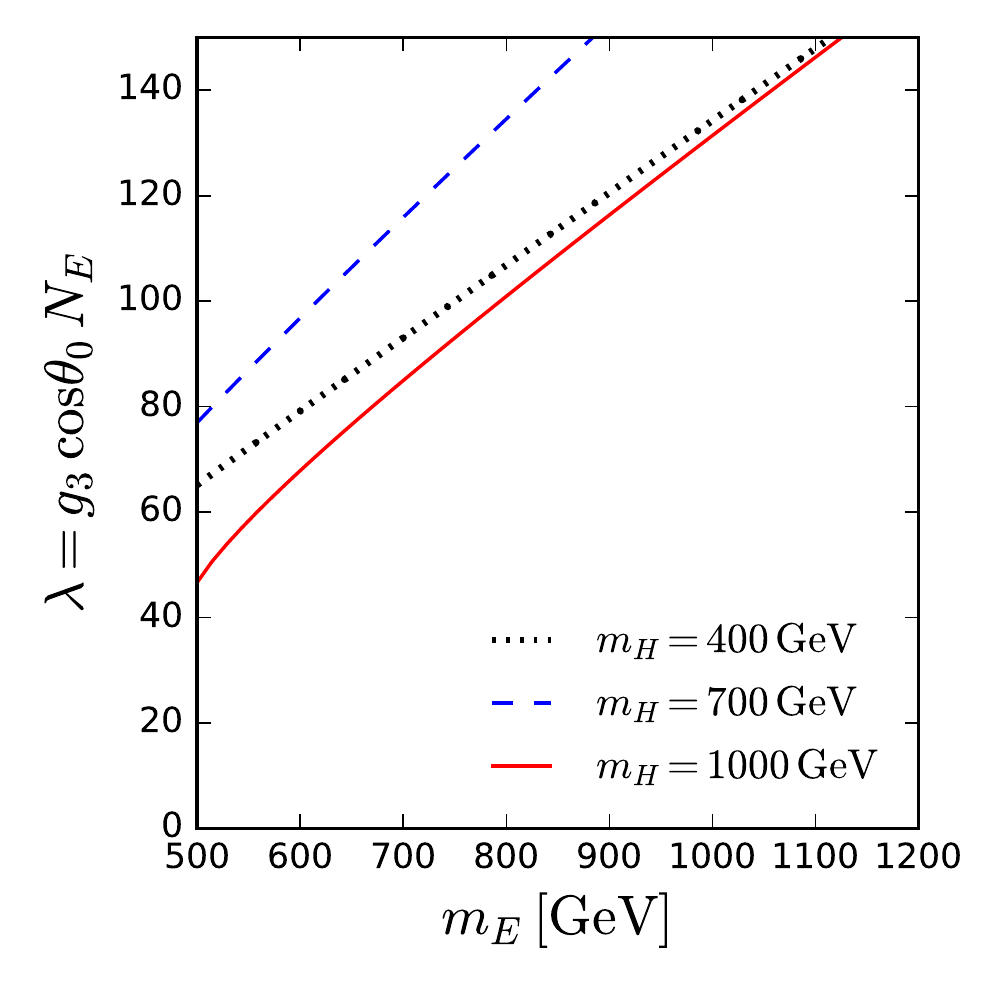}}
\caption{\small Enhancement factor $R_{\gamma\gamma}$ for the $h \rightarrow \gamma \gamma$ decay width in dependence on scalar coupling $c_S$ and the
mass $m_S$ of the lighter charged scalar (left). Bounds allowed by the
LHC 13 TeV search for $pp\to H\to\gamma\gamma$ (right).
}
\label{rgamma}
\end{figure}

{\em The light Higgs} $h(125)\simeq \phi^0\cos\theta_0$ 
is predominantly given by the neutral component of the SM Higgs doublet $\phi^0$, which couples to charged scalars $S=(h^+,\Delta^+)$ via $c_S v \phi^0 S^\dagger S$ coupling, generating diphoton decay amplitude $h(125)\to \gamma\gamma$ at one-loop. Thereby, the $c_S$ couplings correspond to the couplings  $\lambda_4$/2 and $\lambda_5$/2 in (\ref{potential}). The enhancement factor with respect to the SM decay width
is displayed in the left panel of Fig.~\ref{rgamma}. The horizontal lines in this figure highlight the current bound $R_{\gamma\gamma} = 1.14\pm 0.19$~\cite{Khachatryan:2016vau}.
Since the contribution of the lighter among the two charged scalars $S$ dominates, this figure sets a bound on the respective coupling $c_S$. 
Only the charged scalars which are sufficiently light may produce significant effects in the LHC diphoton Higgs signals, so that there is a poor constraint  on $c_S$ couplings of the charged scalars with mass bigger than $\sim$ 400 GeV.
Also, due to the significant asymmetry of the constraint on the negative and positive values of the quartics, {\it negative} $\sim\mathcal{O}(10)$ values of couplings $\lambda_4$ and $\lambda_5$ 
may be excluded only if future considerably larger accumulated statistics at the LHC,
drastically shrinks the bound on $R_{\gamma\gamma}$.
At same time the LHC already starts to probe the perturbative region for the {\it positive} values of the quartics. 
Of course, for the $\sim\mathcal{O}(10)$ values of the couplings one starts to worry about perturbativity of the model and the perturbative bounds on the scalar quartics, analogous to those presented in Ref.~\cite{Khan:2016sxm}. 
As it will be explicated in Sec. 4,  we do not delve into additional assumptions which might enable to deduce some bounds on the quartic couplings, since these anyway do not test heavy (neutral) Higgs. Accordingly, we do not consider the related quartics further in the present short paper. 
The considerations of the  $h(125)\to \gamma\gamma$ here update previous study in~\cite{Brdar:2013iea} done in hope to probe an eventual  charged scalar companion of the neutral heavy Higgs.

In contrast, the quartic couplings in the three-loop model will be essential for the testability of the heavy partner $H$ and their perturbativity bounds will be
considered in Sec.4. Before we move to the three-loop model, we still need to discuss how the scalar $H$ will be experimentally tested in the one-loop model.
\\
{\em The heavy Higgs} $H\simeq \Delta^0\cos\theta_0 $ is predominantly $\Delta^0$ in this model. Let us first discuss the productions mechanisms for $H$. In general $g_{Hff}$ and $g_{HVV}$ are $\sim \sin\theta_0\sim \frac{\langle \Delta^0\rangle}{v}$ which is small. Accordingly,  there is no tree-level coupling of $H$ to SM fermions and vector bosons and the gluon fusion production is negligible. However, if $2m_{\Delta^+}^2=m^2_{H}+m^2_{h(125)}$ the mixing can become sizeable \cite{FileviezPerez:2008bj}, and in this small corner of the parameter space the cascade decay proposed in~\cite{Dermisek:2016via} may become relevant. For special choice of masses specified 
in ~\cite{Dermisek:2015hue}
($m_H<2m_t$  to avoid $H\rightarrow t \bar t$, and
$m_E>m_H/2$   to avoid decays into pair of vector-like leptons)
the cascade decay $H\rightarrow E^\pm+ \ell^\mp  \rightarrow h \ell^+ \ell^- $
can be a dominant discovery mode. 

In the rest of the model parameter space  we are led to consider the EW vector boson fusion (VBF) mechanisms. 
For heavy Higgs much heavier than the electroweak scale, the photon fusion production
mechanism dominates and we neglect the contributions from fusion of
weak bosons (see discussion in \cite{Fichet:2015vvy}). Having in mind the
discovery potential of the Higgs diboson decay, we consider the diboson
signal strength at $\sqrt{s}=13$ TeV from the photon fusion, given by
\cite{Harland-Lang:2016qjy}:
\begin{eqnarray}
\sigma_{VV}\equiv\sigma (pp \to H\to VV )=
\frac{8\pi^2}{m_H} \left.\frac{d\mathcal{L}^{\rm inc}_{\gamma\gamma}}{dM^2}\right|_{M=m_H} 
\!\!\Gamma(H\to\gamma\gamma)\;  {\rm Br}(H\to VV)\;,
\label{gammafusion1}
\end{eqnarray}
where relevant values of inclusive photon-photon luminosity 
$d\mathcal{L}^{\rm inc}_{\gamma\gamma}/dM^2$ for the production
of particle with mass M can be found in \cite{Harland-Lang:2016qjy}.
As we already discussed above, the contributions of the charged scalars to the one-loop-generated $H\gamma \gamma$ coupling are negligible 
and we therefore consider the charged fermion loops with the leading contribution from the Yukawa term
$g_3 \bar{E}_L \Delta E_R+h.c.$ . 
The contributions to different channels with SM gauge bosons for vector-like fermion $E_{L,R}$ with multiplicity $N_E=3$ and coupling $\lambda=g_3 \cos\theta_0 N_E$ read:
\begin{align}
g_{\left( HZ\gamma,\, HZZ \right)} &=  \lambda\alpha\sum_F  \frac{A_{1/2}(\tau_F)}{m_F} \left( \sqrt{2} Q_F\frac{(T_{3F}-s_W^2
Q_F)}{s_W c_W},\; \frac{(T_{3F}-s_W^2 Q_F)^2}{s_W^2 c_W^2}  \right) \,, \nonumber \\
g_{\left( H\gamma\gamma,\, HWW \right)} &= \lambda \alpha\sum_F  \frac{A_{1/2}(\tau_F)}{m_F} 
\left(  Q_F^2,\; \sqrt{2}
\frac{(T_F-T_{3F})(T_F+T_{3F}+1)}{2 s_W^2} \right) \,.
\label{effcoupl}
\end{align}
Here, $T_F$ is the weak isospin of the loop-fermion $F$, the triangle loop function is given by $A_{1/2}(\tau_F)= 2\tau_F (1+ (1-\tau_F )\arcsin^2(1/\sqrt{\tau_F}))$, and the respective variable is $\tau_F=4 m_F^2/m_H^2$. The couplings include
symmetrization factors for identical particles in the final state, and are
normalized so that, neglecting masses of the $W$ and $Z$ bosons, give:
\begin{eqnarray}
\label{finalphotonF}
\Gamma(H \rightarrow VV)=\frac{m_H}{64\pi^3}\bigg|\frac{m_H\, g_{HVV}}{2}\bigg|^2 \ . 
\end{eqnarray} 
For degenerate loop masses, the couplings can be compactly expressed in terms
of quadratic Dynkin indices $I_1$ and $I_2$ of the loop-fermion SM group representations:
\begin{align}
g_{\left(HZ\gamma,\, HZZ \right)} & = \lambda\alpha \frac{A_{1/2}(\tau_F)}{m_F} \left( \sqrt{2}\left(\frac{c_W}{s_W}I_2 - \frac{s_W}{c_W}I_1\right),\; \left(\frac{c_{W}^2}{s_{W}^2}I_2 + \frac{s_{W}^2}{c_{W}^2}I_1\right) \right), \nonumber \\ 
g_{\left( H\gamma\gamma,\, HWW\right)} & = \lambda\alpha \frac{A_{1/2}(\tau_F)}{m_F} \left( \left(I_1+I_2 \right),\; \sqrt{2} \frac{I_2}{s_W^2}\right) \,.
\label{effcouplD}
\end{align}
For the multiplicity $N_E=3$ and $I_1=1/2$, $I_2=1/2$ we obtain for the resulting ratio of the decay widths $R_{VV} \equiv \Gamma(H\to VV) / \Gamma(H\to \gamma\gamma)$
\begin{eqnarray}
R_{WW} \approx 9.1 \,, \quad
R_{ZZ} \approx 3.2 \,,  \quad
R_{Z\gamma} \approx 0.8 \,.
\end{eqnarray}
This results in a branching ratio $Br(H\to \gamma\gamma) \approx 7\,\% $. 

We can now investigate the influence of the constraints coming from the
searches for resonances decaying to gauge boson pairs at the LHC 13 TeV run.
Most stringent constraint comes from searches in diphoton channel, 
where limits on the $\sigma_{\gamma\gamma}$, corresponding to an integrated
luminosity of 15.4 fb$^{-1}$, recorded in 2015 and 2016, can be found in
\cite{ATLAS_15fb}.

The results summarized in Fig.\ref{rgamma}(right) 
show that  for a fixed value of $m_E$, the upper bound on the coupling
$\lambda=g_3 \cos\theta_0 N_E$ generally decreases as we increase the $m_H$ mass,
but upward fluctuation of present LHC data around 700--800 GeV is visible,
making the $m_H=700\,{\rm GeV}$ scenario least constrained.
Numerically, for the multiplicity $N_E=3$, the bound $g_3 \cos\theta_0 <
\mathcal{O}(10-60)$ means that LHC currently starts to constrain the region
where our model is strongly-coupled. For the planned integrated luminosity of 100
fb$^{-1}$, we estimate that LHC will probe $g_3 \cos\theta_0 <
\mathcal{O}(4-20)$ values.

{\bf The three-loop neutrino mass model:}
The CP-even neutral state $H$ in this model emerges from the 2HDM and the contributions to its decays arise from loop-diagrams
with exotic charged scalar particles contained in $\Phi$ and $\chi$ multiplets. 
The quartic vertices which generate these triangle loops 
can be read from two scalar potentials contained in~(\ref{scalarpot}):
\begin{eqnarray}\label{scalarpot-chi-Phi}
  V_m(H_{\bf{1}},H_{\bf{2}},\chi) &\supset& (\tau_1H_{\bf{1}}^\dagger H_{\bf{1}} 
  +  \tau_2 H_{\bf{2}}^\dagger H_{\bf{2}}) \chi^\dagger \chi \ ,  \\
 \nonumber 
V_m(H_{\bf{1}},H_{\bf{2}},\Phi) &\supset&  (\sigma_1H_{\bf{1}}^\dagger H_{\bf{1}}+\sigma_2H_{\bf{2}}^\dagger H_{\bf{2}}) \Phi^\dagger \Phi +  (\sigma'_1H_{\bf{1}}^* H_{\bf{1}}+ \sigma'_2H_{\bf{2}}^* H_{\bf{2}}) \Phi^* \Phi . 
\end{eqnarray}
For scalar septuplet $\chi$ the strengths of trilinear couplings   $h(125)\chi^\dagger \chi$ and $H\chi^\dagger \chi$
are extracted after substituting  the VEVs $v_1=v\  \mathrm{cos}\beta$ and $ v_2=v \ \mathrm{sin}\beta$ in one of the doublets. 
This leads to the couplings of the scalar septuplet $\chi$
\begin{eqnarray}\label{HHchi-tau1,2-vertex}
&V_{\chi}&= (\tau_1 H^{0}_{\bf{1}} H^{0}_{\bf{1}} + \tau_2 H^{0}_{\bf{2}} H^{0}_{\bf{2}})\chi^\dagger \chi  =  \\
     \nonumber        &=&v\chi^\dagger \chi \bigg[H (\tau_1 \mathrm{cos}\alpha \ \mathrm{cos}\beta
+ \tau_2  \mathrm{sin}\alpha\ \mathrm{sin}\beta )+h (- \tau_1 \mathrm{sin}\alpha\ \mathrm{cos}\beta
+ \tau_2  \mathrm{cos}\alpha \ \mathrm{sin}\beta ) \bigg]\ .
\end{eqnarray}
Adopting  the ``alignment limit" condition ($\mathrm{tan}\beta = 1   \ , \   \mathrm{sin}(\beta-\alpha) = 1$)~\cite{Han:2015qqj} for the 2HDM and assuming for the couplings the relation $\tau_1 = -\tau_2\equiv \tau$,  
results in the universal coupling
\begin{eqnarray}\label{HHchi-tilde-tau-vertex}
 V_{\chi}=  \ v \  
\tau \ \big[\mathrm{cos}(\beta+\alpha)   \   H -\mathrm{sin}(\beta+\alpha) \   h\big]\ \chi^\dagger \chi=v \ \tau \  H\chi^\dagger \chi.
\end{eqnarray}
The alignment limit identifies the light state as SM-like $h(125)$, such that
its diphoton decay acquires no contribution  from (\ref{HHchi-tilde-tau-vertex}). 

Equivalent conditions ($\sigma_1 = - \sigma_2\equiv \sigma   \ , \   \sigma'_1 = - \sigma'_2\equiv \sigma'$) set on the couplings for the quintuplet $\Phi$ in (\ref{scalarpot-chi-Phi}), ensure that the trilinear couplings of $h(125)$ to the charged components of the quintuplet vanish. The couplings relevant for the $H \rightarrow \gamma \gamma$ decay explicated in
\begin{eqnarray}   V_\Phi =v H  (c_{\Phi^{+}} \Phi^{+*} \Phi^{+} + c_{\Phi^{-}} \Phi^{-*} \Phi^{-}
                  +  c_{\Phi^{--}} \Phi^{--*} \Phi^{--} + c_{\Phi^{---}} \Phi^{---*} \Phi^{---})  \,
\end{eqnarray}
are given by
\begin{eqnarray}\label{couplings-Phi-H}
  c_{\Phi^{+}}= \sigma, \;
  c_{\Phi^{0}}= \sigma + \frac{\sigma'}{4} ,\; 
  c_{\Phi^{-}}= \sigma + \frac{\sigma'}{2} ,\;
  c_{\Phi^{--}}= \sigma + \frac{3\sigma'}{4} ,\; 
  c_{\Phi^{---}}= \sigma +  \sigma' . 
\end{eqnarray}
Note that the EWSB contributions to the masses of different components of the
septuplet $\chi$ (quintuplet $\Phi$) are given by
$v^2(\cos^2\beta - \sin^2\beta)/2$ multiplied by the coupling $\tau$ ($c_{\Phi^{(Q)}}$), respectively,
so that for $\tan\beta=1$ the EWSB contributions vanish.

\begin{figure}[th]
\centerline{\includegraphics[scale=0.68]{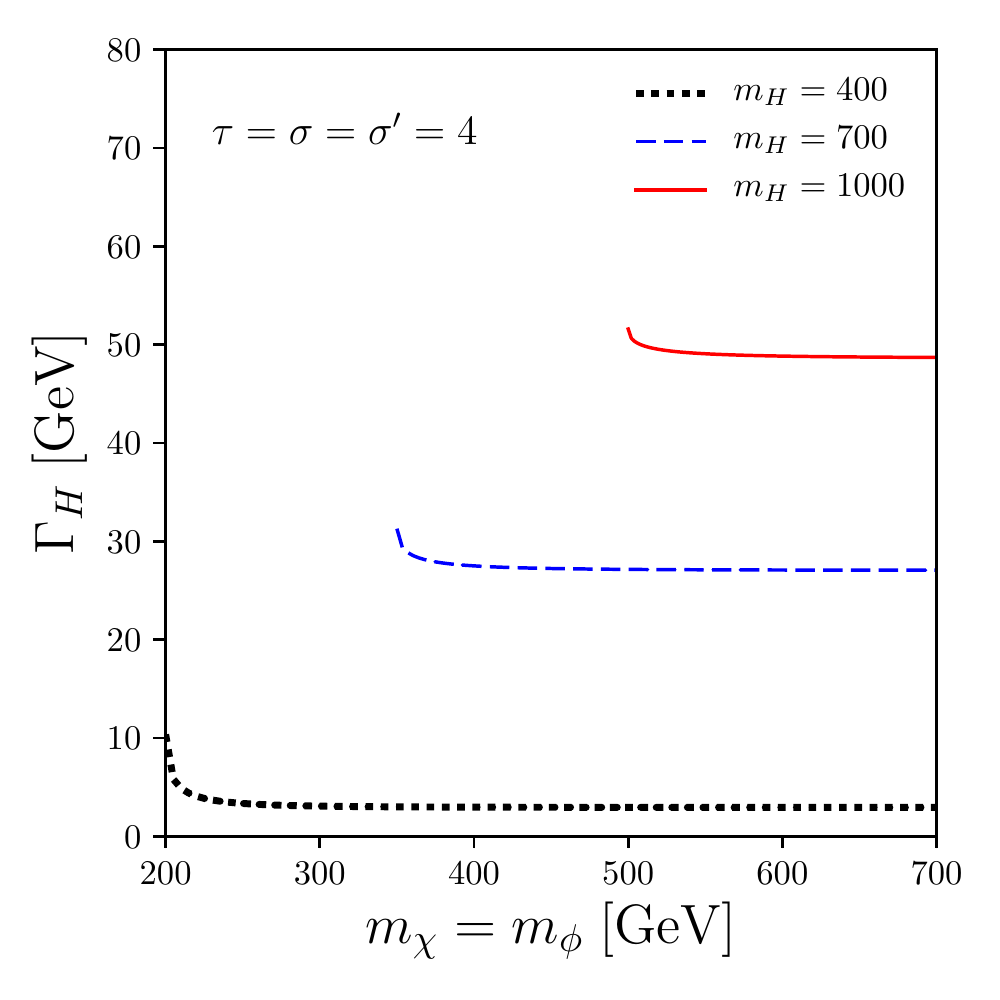}%
\includegraphics[scale=0.68]{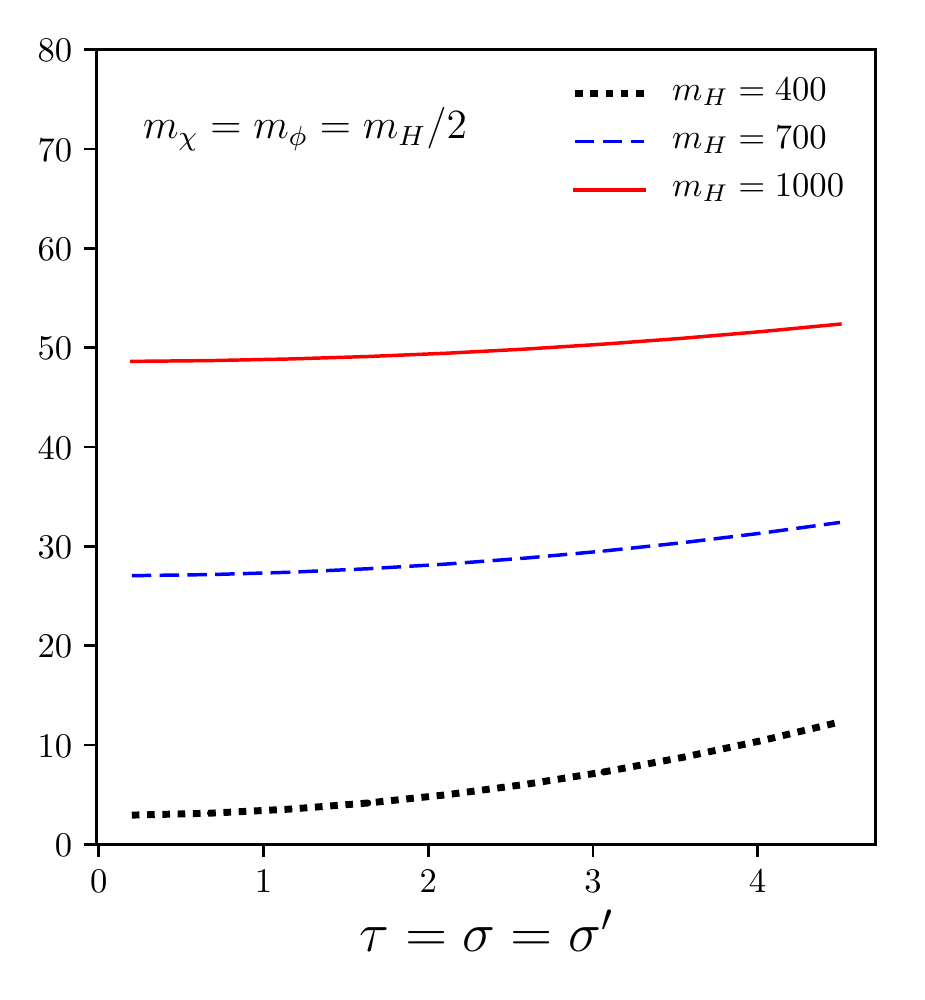}}
\caption{\label{fig:GAMTOT} The total decay width of $H$ particle
in the three-loop neutrino mass model for quartic
couplings in ranges consistent with $m_{\chi,\phi}\ge m_H/2$ region where diboson searches are relevant and with perturbativity bounds deduced in Sec. 4.
}
\end{figure}

\begin{figure}[th]
\centerline{\includegraphics[scale=0.64]{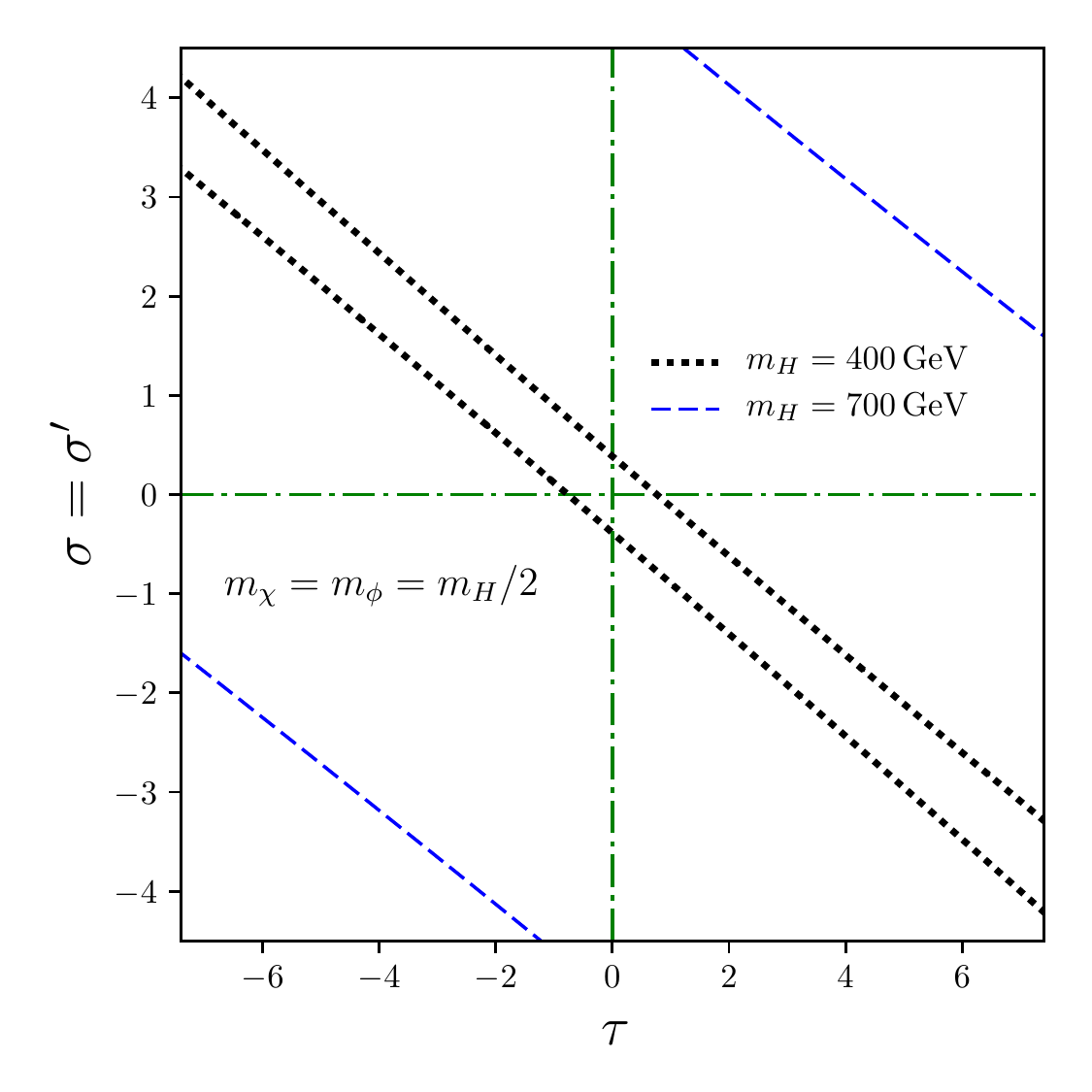}%
\includegraphics[scale=0.64]{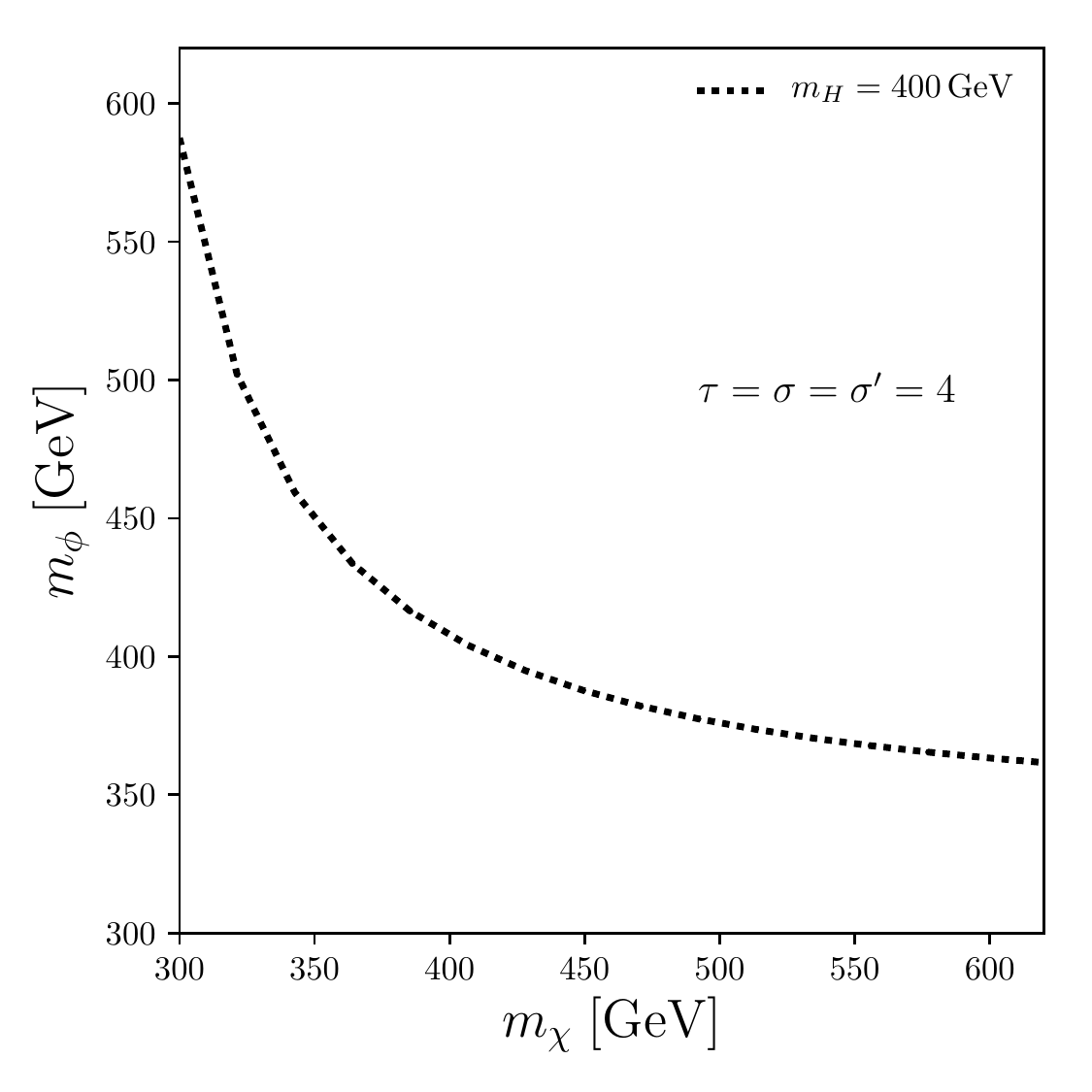}}
\caption{\label{fig:param-ranges} 
Bounds on ranges of values for
coupling (left) and mass (right) parameters
of the three-loop neutrino mass model coming from
ATLAS searches for $p p \to H \to \gamma\gamma$ at $\sqrt{s}=$ 13 TeV \protect\cite{ATLAS_15fb}.
(Parameters for $m_H=$ 1000 GeV (left) and 700, 1000 GeV (right) scenarios are not constrained by diboson searches in the plotted region.)
}
\end{figure}
The quintuplet and septuplet charged scalar states in the loop contribute only
to the decay of the CP-even  boson $H$.
In the lepton-specific 2HDM at hand, only $H_2$ couples to the SM quarks and the relevant couplings of $H$ to
top and $V=W^\pm, Z$ are given in the alignment limit by~\cite{Bertuzzo:2016fmv}
${g_{Htt}}/{g_{Htt}^{SM}} = \cos(\beta-\alpha)-{\sin(\beta-\alpha)}/{\tan\beta}=-1$ and
$g_{HVV} =  2 \cos(\beta-\alpha){m_V^2}/{v}=0$. 
In the scenario where $H$ is produced dominantly through
gluon-gluon fusion, the diphoton cross section is
\begin{equation}
\sigma_{\gamma\gamma}  = \sigma_{ggF}\times Br(H\to\gamma\gamma)\;,
\end{equation}
where the cross section for $pp\to gg X \to H X$ at $\sqrt{s} = 13\,{\rm TeV}$
is $\sigma_{ggF} = 9.66\,{\rm pb}$ for $m_H = 400\,{\rm GeV}$,
$\sigma_{ggF} = 1.03\,{\rm pb}$ for $m_H = 700\,{\rm GeV}$ and
$\sigma_{ggF} = 157\,{\rm fb}$ for $m_H = 1000\,{\rm GeV}$
 \cite{Heinemeyer:2013tqa}.

The tree-level decay width of the scalar $H$ to the $\bar{t}t$ pair is:
\begin{equation}
\Gamma(H\to t\bar{t})= N_c  \frac{\alpha m_H}{8 \sin^2\theta_W} \frac{m_t^2}{m_W^2} \bigg(1-\frac{4 m_t^2}{m_H^2}\bigg)^{3/2}  \,,
\label{eq:Htt}
\end{equation}
\\
and additional contributions to the $H$ width are provided by the decays into SM gauge bosons. These contributions are subleading since the tree-level couplings $H VV$ are absent in the alignment limit, so that these decay modes are generated only at one-loop level. 
To parametrize these decay modes, it is convenient to introduce the effective couplings $g_{HVV}$ of $H$ to the
SM gauge bosons. They can be obtained by substituting 
$\lambda\sum_F \left\lbrace  \cdots\right\rbrace {A_{1/2}(\tau_F)}/{m_F}$ in~(\ref{effcoupl}) with
\begin{equation}
    \tau   \sum_S \left\lbrace \cdots\right\rbrace \frac{v A_{0}(\tau_S)}{2 m_{S}^2} \;\;{\rm and}
    \;\;
\sum_S \left(\sigma + \sigma'\Big(\frac{2-T_{3S}}{4}\Big)\right) \left\lbrace \cdots \right\rbrace
\frac{v A_{0}(\tau_S)}{m_{S}^2} \;,
\label{effcouplS5}
\end{equation}
for the scalar septuplet and quintuplet, respectively. Here the factor  $(2-T_{3S})/4$ accounts for
the non-universality of coupling to $H$ in (\ref{couplings-Phi-H}),
and should be changed to $(3-T_{3S})/8$ in the sole case of $g_{HWW}$.
These constants are normalized so that, neglecting masses of the $W$ and $Z$ bosons, we have
\begin{eqnarray}
\label{finalphotonS}
\Gamma(H\to VV)=\frac{m_H}{256\pi^3}\bigg|\frac{m_H\, g_{HVV}}{2}\bigg|^2 \ . 
\end{eqnarray}
The variable $\tau_S\equiv 4 m_S^2/m_H^2$ and the loop function is given by
$A_0(\tau_S)\equiv-\tau_S (1-\tau_S \arcsin^2 (1/\sqrt{\tau_S}))$.
For the degenerate couplings $\tau=\sigma=\sigma'$, 
this leads to the ratios of diboson to diphoton decay widths 
\begin{eqnarray}
R_{WW} \approx 17.8 \,, \quad
R_{ZZ} \approx 4.9 \,,  \quad
R_{Z\gamma} \approx 3.1 \, .
\end{eqnarray}
The dominance  of the $WW$ channel above can be understood as the quintuplet contributes to both $H\to W^+ W^-$ and $ZZ$ channels, while septuplet, as a real multiplet, contributes only to $H\to W^+ W^-$. 
The total width of $H$ for the generic choice of the parameters is shown in Fig.~\ref{fig:GAMTOT}. It is dominated by the $\bar{t}t$ channel, so that even in the extreme
case when $\tau=\sigma=\sigma'=4$ and for the kinematically most-favored values $m_\chi = m_\Phi = m_H/2$, the branching ratio for the diphoton channel is only
\begin{equation}
 Br(H_{400}\to\gamma\gamma) = 2.7\% \,,\;
 Br(H_{700}\to\gamma\gamma) = 0.5\% \,,\;
 Br(H_{1000}\to\gamma\gamma) = 0.2\% \,.
 \label{eq:BrggCKP}
\end{equation}

As in the one-loop model, we can now investigate the influence of the constraints coming from the
searches for resonances decaying to gauge boson pairs at the LHC 13 TeV run where the most stringent constraint is still coming from the searches in the diphoton channel. Using the limits
on the $\sigma_{\gamma\gamma}$ in \cite{ATLAS_15fb} gives the results shown in Fig.~\ref{fig:param-ranges}. Here (in the left figure) only the strips between the upper and the lower lines, corresponding to the masses $m_H=(400,700,1000)$ GeV, are allowed by the LHC searches, while (in the right figure) the areas below the curves corresponding to the masses $m_H=(400,700)$ GeV are excluded by the LHC searches. 
We note that, owing to larger number of BSM particles in the three-loop model, 
the constraints are stronger than for the one-loop
model. As the mass of $H$ decreases, its couplings to septuplet and quintuplet scalars 
become more (anti)correlated. Also, the Fig.~\ref{fig:param-ranges} (right) shows that, for
the choice of couplings made there, all BSM scalar particles ($H$, $\chi$, $\phi$)
have to be heavier than $\sim 300\,{\rm GeV}$.

\section{Vacuum stability and perturbativity}\label{perturbativity}

A summary of the detailed outcome of the minimal one-loop scenario with a Higgs partner from extra adjoint representation is presented in the first row in Table~\ref{Final}. It is contrasted to a more baroque three-loop model based on a Higgs partner from 2HDM sector in the second row in Table~\ref{Final}.

\begin{table}[h!]
\footnotesize
$$\begin{array}{lc|c|c|c||c|c|c|c|c|}
\hbox{Model}&\hbox{$J^{CP}_{H}$} & \hbox{$\Gamma_{H}$}&\hbox{Production}  & \hbox{Landau Pole}  &\hbox{Br$_{WW}$}& \hbox{Br$_{\gamma\gamma}$}&\hbox{Br$_{Z\gamma}$}&\hbox{Br$_{ZZ}$}&\hbox{Br$_{t\bar{t}}$}\\
\hline
\rowcolor[cmyk]{0.1,0,0.1,0}
\multicolumn{1}{c}{\cellcolor[cmyk]{0,0,0.2,0} \hbox{1-loop}}  &0^{++} &\hbox{3 GeV} & \hbox{$\gamma\gamma$-fusion} &\hbox{Absent} &65\%\ & 7\%\ &6\%\ &23\%\ &- \\
\rowcolor[cmyk]{0.2,0.0,0,0.0}
\multicolumn{1}{c}{\cellcolor[cmyk]{0,0.2,0,0} \hbox{3-loop}} &0^{++}&\hbox{31 GeV}&\hbox{$gg$-fusion}& 10^6 \ \hbox{GeV} &9\%\ & 0.5\%\ &1.6\%\ &2.5\%\ &86\%\ \\
\end{array}
$$
\caption{ Comparison between the neutrino mass models for $m_H=700$ GeV. In the three-loop model the branching ratios are calculated for the benchmark point in (\ref{eq:BrggCKP}).
}
\label{Final}
\end{table}

It is in order to address a UV behavior of these two complementary scenarios. 
Notably, the three-loop model is under a well known threat that invoking large multiplets~\cite{Cirelli:2005uq} leads to Landau poles (LP) considerably below the Planck scale, potentially sensitive to two-loop RGE~\cite{DiLuzio:2015oha} effects.
For the $SU(2)_L$ gauge coupling $g_2$, this threat has been addressed in~\cite{Sierra:2016qfa} for the particle content of two scotogenic three-loop neutrino mass models~\cite{Culjak:2015qja,Ahriche:2015wha} which aim at accidental DM-protecting $Z_2$ symmetry. Thereby the three-loop model at hand~\cite{Culjak:2015qja} is less affected by this threat, and its  exposure to additional scrutiny presented in  \cite{Antipin:2016awv} shows that the LP appears around $10^6$ GeV.

As for the quartic couplings, the large values of the ``mixed" scalar couplings 
$\tau_{1,2}$ and $\sigma_{1,2}^{(\prime)}$ required for potentially observable diphoton signal, and the negative values for some of them 
endanger the stability of the scalar potential and perturbative control over the model. Here, we highlight different remedies that can be envisioned in the septuplet $\chi$ and the quintuplet $\Phi$ quartic sectors, by activating them one at a time.

In the quartic coupling sectors, there are three additional quartic self-couplings of the $\Phi^4$-type and two additional quartic self-couplings of the $\chi^4$-type \cite{Hamada:2015bra} which we are still free to choose. There are additional quartics of the $\chi^2 \Phi^2$-type which we choose to be zero in order to decouple the septuplet and the quintuplet quartic sectors. 

Now, the stability of the potential will be endangered only due to those active ``mixed" quartics which are negative by the virtue of the relations between couplings discussed below (\ref{HHchi-tau1,2-vertex}) and (\ref{HHchi-tilde-tau-vertex}), which may lead to an unbounded potential. Such quartics have to be balanced in the stability condition by appropriately chosen positive values of the corresponding quartic self-couplings (the stability condition for the septuplet sector has been explicated in \cite{Cai:2015kpa}). For the other inactive ``mixed" quartics we may choose the ``self" quartics to be zero at the threshold as well. 

As for the perturbative control of the model, it was shown in \cite{Hamada:2015bra} that for the inactive sector the LP will appear at:
\begin{eqnarray}
    \Lambda_{\Phi}\sim 10^9 \bigg(\frac{m_\Phi}{100 \  \hbox{GeV}} \bigg)^{1.28}\,{\rm GeV}\;, \qquad \Lambda_{\chi}\sim 10^6 \bigg(\frac{m_\chi}{100\  \hbox{GeV}}\bigg)^{1.13} \,{\rm GeV}\;,
\end{eqnarray}
for the quintuplet and the septuplet sectors, respectively. These values are not lower than $10^6$ GeV LP of the mentioned $SU(2)_L$ gauge coupling, so that we have a control over the inactive sector. As for the active scalar, we need to consider the possible Yukawa couplings of this scalar which provide a negative contribution to the one-loop beta function of the quartic self-couplings and may help to push up the LP. Unfortunately, for symmetry reasons, for the septuplet $\chi$ the obvious $\chi \Sigma\Sigma$ choice for the Yukawa term vanishes. Following \cite{Cai:2015kpa}, one may introduce the additional $SU(2)_L$-triplet fermion $\zeta=(3,0)$ to have a Yukawa coupling $\chi \Sigma\zeta$ which may be fine-tuned to delay the appearance of the LP. For the quintuplet $\Phi$, the needed Yukawa coupling $g_{i\alpha}$ already exists in our model~\cite{Culjak:2015qja} 
and can be fine-tuned similarly.

Finally, the dominant contribution to the 1-loop beta functions of the ``mixed" quartics $\tau_{1,2}$ and $\sigma_{1,2}^{(\prime)}$ is given in \cite{Hamada:2015bra}:
\begin{equation}
\beta_{x} \sim 4 x^2 - \frac{153}{2} x g_2^2 + 36 g_2^4 \;, \qquad \beta_y \sim 4 y^2 - \frac{81}{2} y g_2^2 + 18 g_2^4 \;.
\label{RGrun}
\end{equation}

Here, these couplings are denoted by $x=\tau_{1,2}$ and $y=\sigma_{1,2}^{(\prime)}$ collectively. Due to the large negative coefficients in $g_2^2$ terms, it is easy to check that for $x$ $(y)$ in the regions $x<0$ $(y<0)$ and $0.2<x<7.9$ $(0.2<y<4.1)$ the sign of the  respective beta function is such that by the running of the ``mixed" quartic coupling its initial value will be driven towards decreasing its absolute value\footnote{We took the SM value of the $SU(2)_L$ gauge coupling $g_2 (100 \ \hbox{GeV})\approx 0.65$. }. Combined with the initial conditions $\tau_1=-\tau_2$ and $\sigma_1^{(\prime)}=-\sigma_2^{(\prime)}$ discussed below (\ref{HHchi-tau1,2-vertex}) and (\ref{HHchi-tilde-tau-vertex}) respectively, this translates into the bounds $0.2<|x|<7.9$ and $0.2<|y|<4.1$. As we increase further the energy, the $SU(2)_L$ gauge coupling $g_2$ increases towards its LP and the $g_2^4$-term will eventually start to dominate the evolution, driving these ``mixed" quartics to the LP as well. We therefore expect that the dangerously-large initial values of the ``mixed" quartics needed for observable signal will develop LP $\sim 10^6$ GeV together with the $g_2$ coupling.

For completeness, let us address also  the running of the quartic couplings in the one-loop model, using $\lambda_5$ as an illustrative example. 
The absence of Landau pole for SM couplings in case of the one-loop model does not justify the truncation to the same terms used in (\ref{RGrun}) and leads us to consider additional terms in the beta function. These terms include in addition to
all leading SM terms from \cite{Khan:2016sxm} also 
other BSM quartics in one-loop model, so that we obtain
\begin{equation}
\beta_{\lambda_5} \sim 8 \lambda_5^2 - \frac{33}{2} \lambda_5 g_2^2 + \frac{3}{2} g_2^4 -\frac{9}{10} \lambda_5 g_1^2+ 6 \lambda_5 y_t^2+ 40 \lambda_3 \lambda_5+12\lambda_1 \lambda_5+ a \lambda_6 \lambda_5\;.
\end{equation}
Here, $\lambda_1$ is the quartic of the SM Higgs and the $a\sim \mathcal{O}(1)$ coefficient is immaterial for our conclusion. 
This beta function does not allow us to set perturbativity  bounds, since 
 $\lambda_{3}$ and  $\lambda_{6}$ quartics are not constrained by $H\gamma\gamma$ vertex anyway, as they vanish in the $\langle \Delta_0\rangle=0$ limit.

\section{Conclusions}\label{Conclusions}

The discovery of a Higgs-like $125$ GeV particle opened a question of its possible relatives.
Additional Higgs bosons employed in attempts to produce neutrino masses radiatively may be accessible at the LHC.
Two selected radiative neutrino mass scenarios~\cite{Brdar:2013iea,Culjak:2015qja} possess accidental protective discrete symmetries in the same way as we expect from viable TeV extensions of the SM to preserve the accidental baryon number of the SM to sufficient accuracy. 
These scenarios came recently under a serious threat raised by a claimed 750 GeV diphoton resonance, addressed in our unpublished paper~\cite{Antipin:2016awv}.
Now, after the hints of 750 GeV resonance have faded away, our focus turns here to testability of our models and their validity with respect to appearance of Landau poles (LP) or an instability of the scalar potential.
While the one-loop model~\cite{Brdar:2013iea} is safe up to the Planck scale, the three-loop model~\cite{Culjak:2015qja} may be under a more serious threat. However, in the latter case 
the mixture of the 2HD and the exotic scalar sector provides a fortuitous remedy for the too early
LP for relevant couplings. Due to signs and sizes of the coefficients in the 
relevant beta functions the LP is relegated beyond the $\sim 10^6$ GeV scale.
Far below this scale the exotic states of three-loop model already enter into the play.

As shown in Section 3, the testability of the one-loop model is enabled by a presence of inherent vector-like leptons. In the very special corner of the parameter space the vector-like leptons mediate the heavy Higgs cascade decay $H\rightarrow E^\pm  + \ell^\mp  \rightarrow h \ell^+ \ell^- $ as potential discovery mode has to be mentioned for completeness.
In the complementary region of one-loop model parameter space the vector-like leptons $ E\sim (2,-1)$
lead to loop-generated diphoton decay as potential discovery channel.
In the case of three-loop model, the vector-like leptons $\Sigma \sim (5,0)$ 
are decoupled from 2HD fields and the diphoton decay, generated by
exotic scalar fields of this model, remains the main discovery mode.
For both models we have compared branching ratios of various diboson
decay channels and presented
constraints imposed by recent LHC diphoton resonance searches.

\subsubsection*{Acknowledgment}
This work is supported by the Croatian Science Foundation under the project number 8799 and by the QuantiXLie Center of Excellence. OA was partially supported by the H2020 CSA Twinning project No.692194, RBI-T-WINNING.

\end{document}